\RequirePackage{silence}
\documentclass{_nature2_modified}

\usepackage[utf8]{inputenc}
\usepackage{color}
\usepackage{xcolor}
\usepackage{amsmath,amssymb}
\usepackage{graphicx}
\usepackage{hyperref}
\usepackage{lineno}
\usepackage[numbers,sort&compress]{natbib}

\ifx\pdfsuppresswarningdupdest\undefined\else\pdfsuppresswarningdupdest=1\fi
\hypersetup{bookmarksdepth=2}
\providecommand{\bibcommenthead}{}

\newcommand{\obj}{J0439+1634}
\newcommand{\kms}{\ensuremath{\mathrm{km~s}^{-1}}}

\title{\flushleft
Discovery of Quasar Variability and Early Accretion Disk Signatures at Cosmic Dawn
}

\author{
Gene C. K. Leung$^{1}$,
Anna-Christina Eilers$^{1,2}$,
Christos Panagiotou$^{1}$,
Julien Wolf$^{3}$,
Kishalay De$^{4,5}$,
Luke Weisenbach$^{6}$,
Minghao Yue$^{7}$,
Xiaohui Fan$^{7}$,
Yuzo Ishikawa$^{1}$,
Erin Kara$^{1,2}$,
Mirko Krumpe$^{8}$,
Andrea Merloni$^{9}$,
Robert A. Simcoe$^{1,2}$,
Feige Wang$^{10}$,
Jinyi Yang$^{10}$
}

\begin{document}
\maketitle

\vspace{-0.8cm}
\begin{affiliations}
\item MIT Kavli Institute for Astrophysics and Space Research, Massachusetts Institute of Technology, Cambridge, 02139, Massachusetts, USA
\item Department of Physics, Massachusetts Institute of Technology, Cambridge, 02139, Massachusetts, USA
\item Max-Planck-Institut f\"ur Astronomie, Heidelberg, 69117, Germany
\item Department of Astronomy and Columbia Astrophysics Laboratory, Columbia University, 550 W 120th St. MC 5246, New York, 10027, New York, USA
\item Center for Computational Astrophysics, Flatiron Institute, 162 5th Ave., New York, NY 10010, USA
\item Institute of Cosmology and Gravitation, University of Portsmouth, Dennis Sciama Building, Burnaby Road, Portsmouth, PO1 3FX, UK
\item Steward Observatory, University of Arizona, 933 North Cherry Avenue, Tucson, 85721, Arizona, USA
\item Leibniz-Institut f\"ur Astrophysik Potsdam (AIP), An der Sternwarte 16, Potsdam, D-14482, Germany
\item Max-Planck-Institut f\"ur Extraterrestrische Physik, Garching, 85748, Germany
\item Department of Astronomy, University of Michigan, 1085 South University Avenue, Ann Arbor, 48109, Michigan, USA
\end{affiliations}

\vspace{0.6cm}
\begin{abstract}
{\boldmath
In the nearby universe, quasars are well known to exhibit variability in their brightness over time, offering a powerful tool to probe the physics of accretion onto the SMBH and directly measure the mass of the SMBH. However, detecting variability in early quasars remains challenging. Here, we report the detection of multi-wavelength infrared and X-ray variability in a quasar observed just 850 million years after the Big Bang. The infrared variability spans five filters, tracing rest-frame ultraviolet and optical emission from the accretion disk, while the X-ray variability probes the corona. The variable spectrum reveals that the accretion disk has a geometrically thin, optically thick structure. This provides observational constraints on the accretion disk structure at early times, when quasars are accreting at high Eddington ratios and reside in extreme environments. Our findings demonstrate the feasibility of characterizing accretion physics using variability in the early universe, laying the groundwork for studies exploiting upcoming facilities such as the Rubin Observatory and Roman Space Telescope. These facilities will discover large samples of variable high-redshift quasars, enabling population-level variability studies of accretion physics and black hole masses, filling key missing ingredients in understanding early SMBH growth.
}
\end{abstract}

\section{Main}\label{sec:main}

Quasar variability encodes key information on its accretion state and has been used to characterize its accretion structures from the local universe \citep{fausnaugh16, cackett20} out to $z=2.66$ \citep{pozonunez25}.
Until now, studies of rest-frame optical quasar variability \citep[e.g.][]{vandenberk04, macleod12, goncalves25} were primarily based on multi-epoch data from ground-based surveys such as the Sloan Digital Sky Survey (SDSS) \citep{york00} or Zwicky Transient Facility (ZTF) \citep{bellm19}, which only covers the rest-frame optical ($>3000$ \AA) up to redshifts of $z \lesssim 2$.
However, high-redshift quasars provide a unique laboratory to probe how accretion physics changes under extreme conditions in the distant universe that are uncommon in local AGNs. Most known quasars at $z \sim 6$ accrete at substantially higher Eddington ratios than their local counterparts \citep{fan23}, where the accretion disk structure may differ from that at lower accretion rates \citep[e.g.][]{abramowicz88, wang99}. The early universe also presents vastly different environments around the quasars, with less massive galaxies at high redshift hosting reduced heavy elements \citep{nakajima23, curti24}, which may strongly influence accretion disk stability and structure \citep[e.g.][]{jiang16}.
The study of quasar variability at high redshift therefore offers a critical, previously unexplored frontier by probing early cosmic time and extreme physical conditions.

\begin{figure}[t]
\centering
\includegraphics[width=\textwidth]{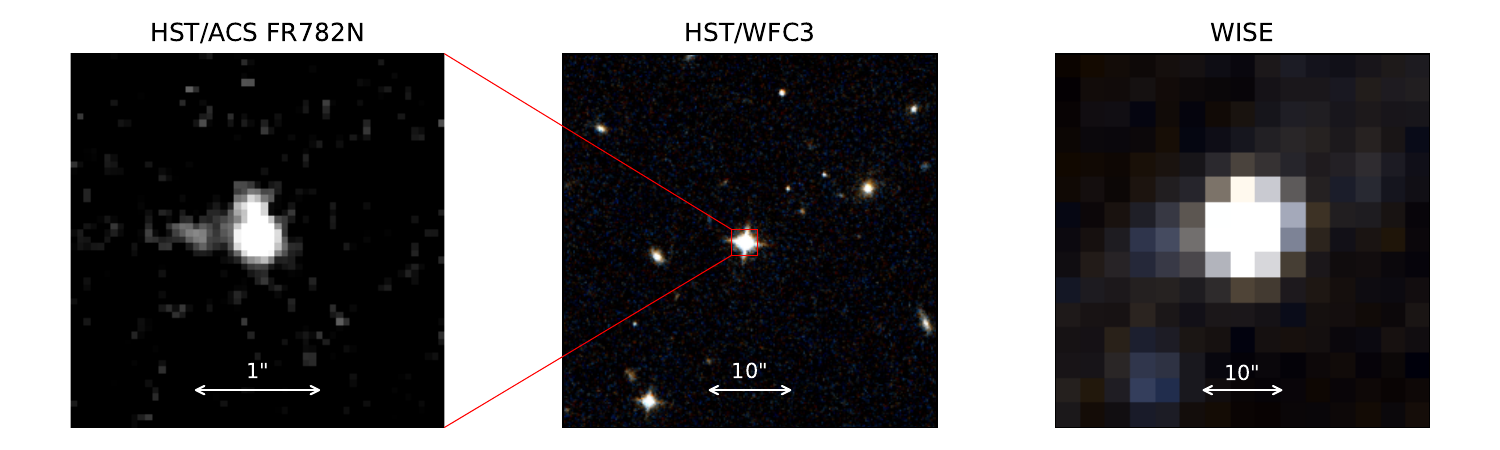}
\caption{{\bf HST and WISE images of \obj .}  Left: $2.75''\times2.75''$ image cutout in the HST/ACS FR782N filter, adapted from \citet{fan19}. The lens galaxy is seen to the left of the quasar images. Middle: $40''\times40''$ color image in the HST/WFC3 F125W and F160W filters. The red square indicates the size of the FR782N cutout. Right: $40''\times40''$ color image in  the WISE W1 and W2 filters from eight-year unWISE coadds \citep{meisner22}.}\label{fig:image}
\end{figure}

The Wide-Field Infrared Survey Explorer (WISE) \citep{wright10} provides a unique capability for studying quasar variability at the highest redshifts. WISE, later reactivated as the NEOWISE mission, repeatedly surveyed the entire sky from 2010 to 2024 in the infrared wavelengths, making it the only facility to provide regular, multi-epoch observations of the rest-frame optical emission of quasars out to $z \sim 7$. We recently conducted a systematic search for high-redshift quasar variability using WISE data, targeting all known quasars at $z > 5.3$ \citep{fan23}. Our calibration procedures are described in Methods (see also Extended Data Fig. \ref{fig:chi2} and \ref{fig:cal}). In this article, we report the discovery of significant variability in the $z=6.51$ quasar J043947.08+163415.7 (hereafter \obj).
This object was first reported in \citet{fan19} as a gravitationally lensed quasar, revealed by multiple images in Hubble Space Telescope (HST) imaging (Fig. \ref{fig:image}). After accounting for a magnification factor of $51.3 \pm 1.4$, it was found to host a black hole of $(6.3 \pm 0.2) \times 10^{8}~M_\odot$ \citep{yang21}. Typical of quasars at these redshifts, it is rapidly accreting close to the Eddington limit, with an Eddington ratio of $(60\pm 10) \%$ \citep{yang21}, substantially higher than average AGNs in the local universe.

\begin{figure}[htbp]
\centering
\includegraphics[width=0.99\textwidth]{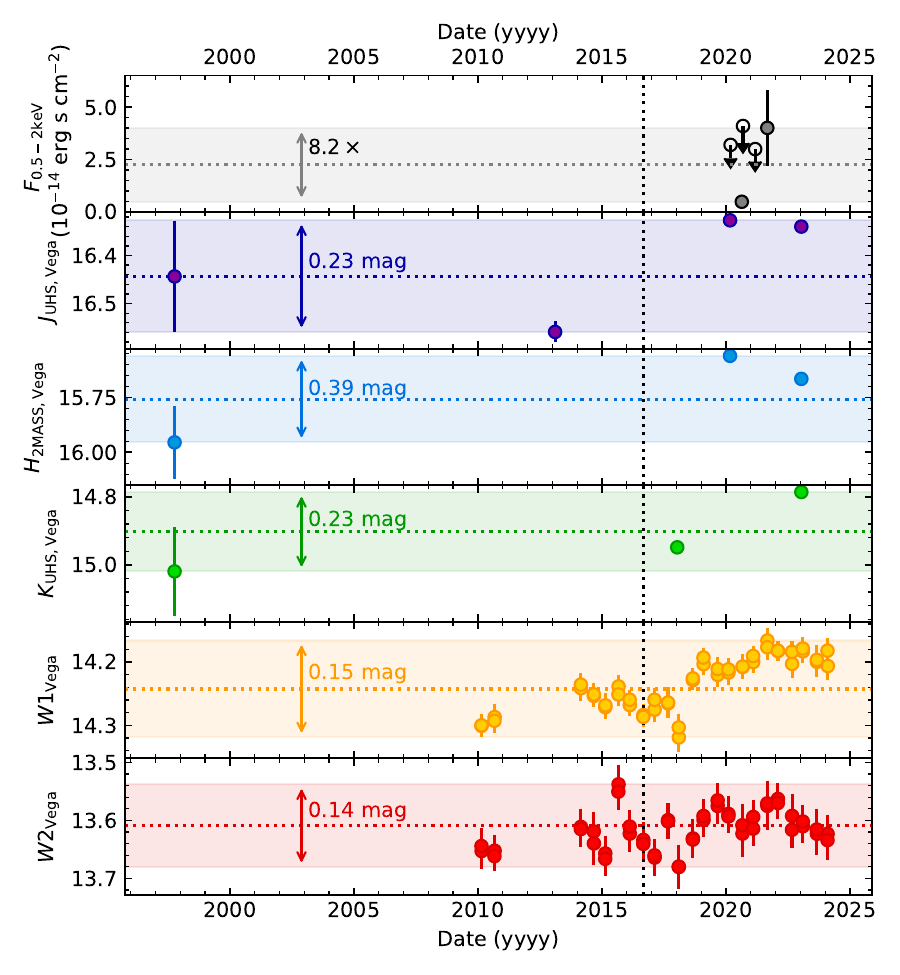}
\caption{{\bf Multi-wavelength light curves of \obj \ from 2000 to 2025.} Data in the 0.5-2 keV, UKIDSS $J$, $H$, $K$ and WISE W1 and W2 bands are shown from top to bottom. Error bars represent the $1\sigma$ uncertainties, and the X-ray upper limits are $1\sigma$. The shaded regions show the range spanning the maximum and minimum magnitudes observed within the duration of the light curves, while the horizontal dotted lines show the mid-point. The flux of \obj \ varied significantly in W1 by $\sim 0.1$ mag at the $10.9 \sigma$ level. The amplitude of the variation is $\sim 0.2$ mag in the bluer $JHK$ bands, and 8 times in the X-ray.}\label{fig:lc}
\end{figure}

\obj \ exhibited variability in both the WISE W1 and W2 filters, corresponding to 3.4 and 4.6 $\mu$m. We show the WISE light curves of J0439 in the bottom two panels of Fig. \ref{fig:lc}. An increase in the quasar's luminosity can be observed in both WISE bands since approximately 2016, reaching a maximum at around 2021. The short duration of the flux increase is inconsistent with a potential microlensing event, which is expected to span $\approx 45$ years at these redshifts (see Methods). The flux difference between the minimum and maximum is 0.15 mag in the W1 filter and 0.14 mag in the W2 filter. We quantify the significance of the variability by calculating the $\chi^2$ of the observed fluxes relative to the median flux. The variability is significant at $10.9 \sigma$ level with $\chi^2=235$ (d.o.f.$=45$) in the W1 filter, and $1.5\sigma$ ($\chi^2=53$, d.o.f.=39) in the W2 filter.

Multi-epoch observations in other IR bandpasses are available for this quasar, as it was observed in the $JHK$ IR filters by 2MASS in 1997 \citep{skrutskie06}, $JK_s$ by UHS in 2013 and 2018 \citep{dye18, schneider25}, F125W and F160W by HST in 2020 (PID 15829, PI: Yang), and the JWST/NIRSpec IFU in 2023 (PID 1222, PI: Willott). This allows us to trace for the first time the multi-band luminosity evolution of a $z>6$ quasar over five different IR filters from $\sim 1$ to 5 $\mu$m, corresponding to UV-optical wavelengths of $\sim 1000-6000$~\AA\ at the rest frame. We show the $JHK$ light curves, transformed to common band passes, in the second to fourth panels of Fig. \ref{fig:lc}. The multi-band photometry shows that, in line with the WISE bands, the $JHK$ fluxes have increased significantly after 2016. In particular, the brightness in the $J$ band has increased by 0.22 mag between 2013 and 2020. These bluer bandpasses show a larger variation amplitude of up to $\sim 0.2$ mag compared to the WISE filters.

\obj \ was also observed in the X-ray by XMM Newton in August 2020 \citep{Yang22}, as well as by eROSITA for four times between 2020 and 2021 in six month intervals. The X-ray light curve is shown in the bottom panel of Fig. \ref{fig:lc}. The 0.2-5 keV X-ray flux of \obj \ has increased by $8.2 \pm 3.7$ times from mid-2020 to late 2021, demonstrating a significantly larger variation amplitude than the infrared bands. Stronger variability in the X-ray compared to the UV and optical is also commonly seen in quasars in the local universe, which has been interpreted as a more compact X-ray-emitting corona than the thermally-emitting accretion disk \citep{kara25}.

We explore the physical origin of the variability of \obj \ by examining the relevant timescales associated with an accretion disk. The brightening in the WISE data spans approximately 1830 days in the observed frame, corresponding to $\sim 240$ days in the rest frame of the quasar after accounting for cosmological time dilation at $z=6.51$. There is also a noticeable flux difference between each epoch, which are $\sim 180$ days apart, or 24 days in the rest frame. In the discussion below, we will adhere to the observed frame time. The shortest relevant timescale to consider is the light-crossing time of the accretion disk. In the lamp-post model \citep{cackett07}, where the accretion disk reacts to varying illumination from a compact corona, the observed variability is expected to follow this timescale. Assuming a standard thin disk, the region of the accretion disk emitting at the rest-frame wavelength corresponding to the W1 filter is $\sim 200 R_S$, where $R_S$ is the Schwarzschild radius. This yields a light crossing timescale of $\sim 100$ days. This is comparable to the short term flux variation between successive epochs.

The orbital timescale, which represents the time over which gas orbital motion occurs,
is $\sim 40$ years for a region of size $200 R_S$ within the accretion disk. This is longer than the brightening of \obj \ from 2016 to 2022, and smaller flux variability can be observed in shorter timescales. 
The thermal timescale, corresponding to the propagation of thermal fluctuations through the disk \citep{kelly09}, is $\sim 200$ years. Similarly, the viscous timescale, corresponding to a change in the global accretion rate through radial inflow \citep{netzer13}, spans $\sim 10^5$ years. These two timescales are exceedingly longer than the observations.
Considering the different timescales, the observed variability in J0439+1634 is most likely associated with processes occurring on light-crossing timescales, such as the reprocessing of emission from a variable high energy source by the accretion disk at longer wavelengths. This is also the prevailing picture for variability observed in nearby AGNs \citep[][]{cackett07, fausnaugh16, cackett20}, showing that the driving mechanism for the variable accretion disk emission in high redshift quasars may be similar to those in the local universe.

\begin{figure}[htbp]
\centering
\includegraphics[width=0.5\textwidth]{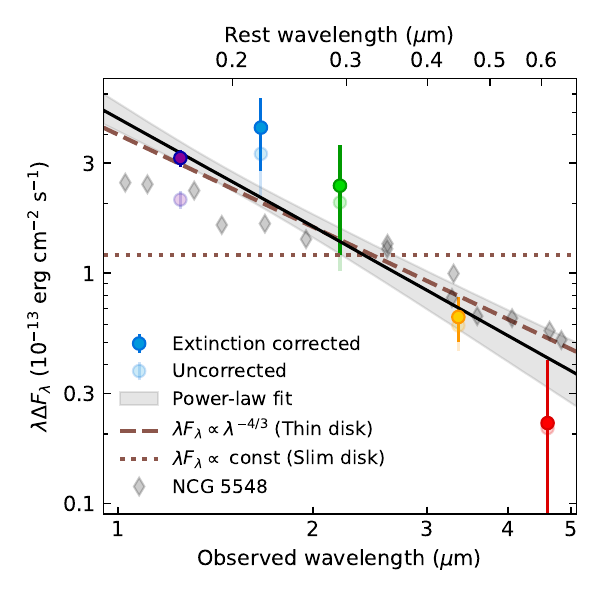}
\caption{{\bf The variable SED of \obj .} The difference between the maximum and minimum fluxes in each filter is plotted against wavelength. The point colors indicate the different filters, using the same color scheme as Fig. \ref{fig:lc}. The dark points are corrected for Milky Way extinction, while the faint points are the uncorrected fluxes.  Error bars represent the $1\sigma$ uncertainties. The black line and grey shaded regions show the best-fit relation and its $68\%$ confidence interval. The brown dashed line shows the expected spectrum from a standard geometrically thin, optically thick accretion disk, while the brown dotted line shows that of a slim accretion disk. For comparison, we also show the variable SED of the local AGN NGC 5548 \citep{fausnaugh16}, corrected for redshift and scaled for visualization, in the grey points. The variable SED J0439+1643 is consistent with the thin disk model, indicating the the flux variation originates from variations in the disk emission.}\label{fig:varspec}
\end{figure}

The long-term variability observed in multiple infrared filters in this quasar allows us to characterize the physical origin of the variability. 
We show the variable spectrum, defined as the difference between the maximum and minimum fluxes as a function of wavelength, in Fig. \ref{fig:varspec}. 
The variable spectrum isolates the variable component of the emission, providing a unique view into the accretion disk spectral energy distribution (SED) \citep[e.g.][]{fausnaugh16, cackett20}.

Theoretical models generally agree that accretion disks are geometrically thin at low accretion rates ($\lesssim 10\%$ of the Eddington limit \citep{shakura73}).
At extremely high accretion rates ($\gg 100\%$ Eddington), radiation pressure is expected to change the structure of the accretion flow, producing in a slim accretion disk \citep{abramowicz88, wang99}.
However, due to scarce observational constraints, there is currently no consensus on the accretion structure in the transitional regime approaching the Eddington limit, which is the dominant mode in the early universe. The high Eddington ratio of $(60 \pm 10)\%$ in \obj \ offers a unique opportunity to constrain theoretical models with observations.

We find that the variable spectrum is well represented by $\lambda f_\lambda \propto \lambda^{-4/3}$, which is expected for a standard geometrically thin, optically thick accretion disk \citep{shakura73}.
The best-fit value of the spectral slope is $-1.58 \pm 0.25$, consistent with the thin disk expectation of $-4/3$. By contrast, a slim disk, which is believed to represent highly accreting systems, is expected to produce a spectrum of $\lambda f_\lambda = \mathrm{constant}$ \citep{wang99}, inconsistent with the observations here. We note that the broad Balmer emission line contribution to the WISE filters are $\lesssim 10 \%$ of the continuum (see Methods and Extended Data Fig. \ref{fig:nirspec}), and any potential variable emission from the broad line region gas is minor in the observed trend.
This shows that even at such high redshift and a high Eddington ratio of $(60 \pm 10)\%$, the accretion disk follows a geometrically thin, optically thick structure.

With the long-term variability monitored by WISE, we estimate the accretion disk size using the time lag between the W2 and W1 light curves of \obj . No significant time delay is observed within the $\sim 180$ days cadence of the WISE observations, with a $90\%$ upper limit of 160 days (see Methods and Extended Data Fig. \ref{fig:iccf}). For a standard thin disk structure, the time lag is expected to scale as $\lambda^{4/3}$ \citep{shakura73, cackett07}, which is commonly observed by reverberation mapping studies of the accretion disk in the local universe \citep[e.g.][]{edelson15, fausnaugh16}. Since the accretion disk spectrum of \obj \ is consistent with a thin disk, we estimate the disk radius using the upper limit on the time lag between W2 and W1, and extrapolating the $\lambda^{4/3}$ relation. This yields a radius upper limit of $<48$ light-days in the rest frame for the portion of the disk where the temperature produces a black body spectrum peaking at the W2 wavelength. The observed constraints are consistent with the thin disk prediction for the estimated black hole mass and Eddington ratio of \obj \ of 22 light-days from the W2 temperature to the ionizing continuum. 

Our results also open the door for new methods for measuring black hole masses at high redshift, which are currently highly uncertain \citep[e.g.][]{abuter24}. 
At the moment, black hole masses at high redshift are commonly derived from scaling relations of emission line widths calibrated in the local universe \citep[e.g.][]{greene05}, the validity of which has so far been unverified at high redshift. Therefore, independent SMBH measurements are crucial in addressing fundamental questions such as the origin of black hole seeds or growth mechanisms to produce $10^9 M_\odot$ SMBHs by $z\sim 6$. \obj \ represents the best target for reverberation mapping (RM) \citep[e.g.][]{kaspi00} to directly measure the mass of a SMBH at such early epochs. While RM for high-redshift quasars has traditionally been difficult due to the long time lag associated with high luminosity at high redshift, this quasar benefits from gravitational lensing magnification, allowing us to probe lower intrinsic luminosity despite its observed brightness, thereby reducing the time needed for RM and thus making this the most accessible target for such a measurement at $z>6$. Using the broad-line region radius-luminosity relation \citep{greene05}, the light crossing time for the broad-line region of \obj \ is $\sim 10$ years in the observed frame, shorter than $\sim 85 \%$ of the $z \gtrsim 5.3$ quasar sample, which typically have light crossing times greater than $\sim 20$ years. Furthermore, simulations have predicted a correlation between the accretion disk and broad-line region sizes \citep{panda24}, which has been tentatively verified by observations \citep{pozonunez25}. This correlation will significantly shorten the required monitoring time by $\sim 100$ times to infer the black hole mass by monitoring the more compact accretion disk instead of the broad-line region. \obj \ represents the best target with known accretion disk variability to perform accretion disk reverberation mapping to infer an independent black hole mass measurement at $z \sim 6$.

In this study, we demonstrate the feasibility of using variability as a tool to characterize the accretion disk structure and obtain independent black hole mass measurements at high redshift. Our work has laid the groundwork for recent and upcoming facilities such as the Roman Space Telescope and Rubin Observatory, which are expected to detect a large number of variable quasars in the early universe. 
In the near future, similar studies of variable quasars will become possible on the population level, providing new insights into the key physics governing the growth and accretion of early SMBHs.

\section{Methods}\label{sec:methods}

\setcounter{figure}{0}
\renewcommand{\thefigure}{\arabic{figure}}
\renewcommand{\figurename}{Extended Data Fig.}
\renewcommand{\theHfigure}{ext.\arabic{figure}}

\subsection{WISE data and analysis}

We conduct a variability search on the compilation of 531 known high-redshift quasars at $z > 5.3$ in \citet{fan23} using images from the unWISE individual-image catalog \citep{schlafly19}. The unWISE images were taken in an approximately 6-month cadence between 2009 and 2024, with each epoch consisting of coadds of typically $12\times7.7$ s exposures in the W1 (3.4 $\mu$m) and W2 (4.6 $\mu$m) channels. For each quasar, we construct WISE light curves in the W1 and W2 bands by performing aperture photometry. A detailed description of the photometry pipeline will be presented in K. De et al. [in prep.], and we briefly describe it below.

We perform simple aperture photometry using a fixed circular aperture with a 2 pixel radius, equivalent to 5.5'', at the sky positions of each quasar in the \citet{fan23} catalog. The background is measured in an annulus with an inner radius of 5 pixels and outer radius of 8 pixels, and is subtracted from the source. The uncertainties are estimated using the uncertainty maps provided by unWISE, which map uncertainties to the pixel level \citep{lang14}.

The resulting WISE light curves typically consist of 20 epochs in a six-month cadence spanning 2009 to 2024, with a gap in the observations in 2011 and 2012. For a small number of quasars that fall within the overlapping region between two WISE pointings, two epochs in close succession are available every six months, resulting in typically 40 epochs in the light curves. From these light curves, we first select quasars with a high confidence detection at a signal-to-noise ratio of greater than 20 in at least 10 epochs in the WISE bands for further analysis and calibration. The high signal-to-noise enables the detection of small amplitude variability, while the number of epochs allows the detection of long-term variability. This results in a sample of 30 quasars for further analysis.

To remove any systematic variation in the photometry over time, we perform flux calibration on the light curves using reference stars within a 0.5$^\circ$ radius in the Gaia DR3 catalog. To ensure the photometric and astrometric quality of the reference stars, we selected sources with (1) a stellar classification probability $>0.9$, (2) no photometric variability flag, (3) neither a quasar or galaxy candidate flag, (4) no non-singular star flag, (5) astrometric excess noise $<$ 2 mas, (6) no neighboring sources within 12'', approximately twice the WISE PSF at W1 and W2. To account for any potential magnitude-dependent photometric offsets, we further require that the magnitudes of the reference stars are within $\pm 0.5$ mag of the quasar in the corresponding WISE filter. This typically results in $\sim 100-200$ reference stars per quasar per epoch.

\begin{figure}[t]
\centering
\includegraphics[width=0.9\textwidth]{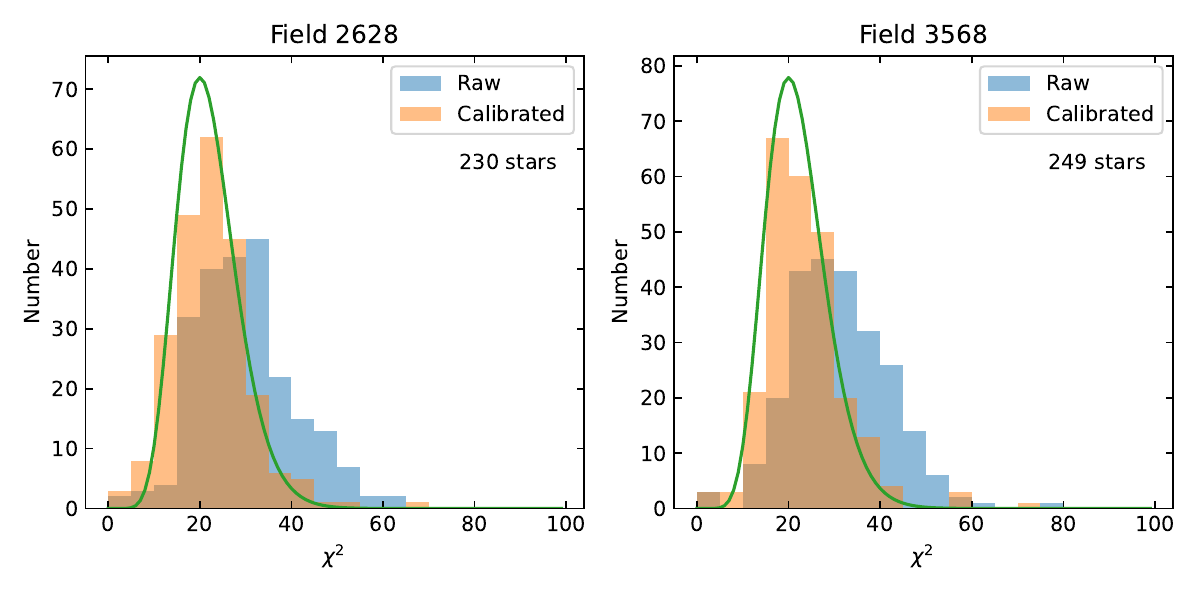}
\caption{{\bf The $\chi^2$ of the reference stars for \obj .} Values before and after flux calibration are shown in the blue and orange histograms, respectively. The two panels show the two WISE pointings in which \obj \ is located. The expected $\chi^2$ distribution with the corresponding degree of freedom is shown in the green line. The flux calibration successfully removes systematic photometric variations in the light curves.}\label{fig:chi2}
\end{figure}

We then calibrate for and remove any systematic photometric offset at different epochs in the light curve. To do this, we normalize the light curves of each reference star by the first epoch, before combining them using a median stack to create a reference light curve. We then subtract this median reference light curve in magnitude space from that of the quasar to create a calibrated light curve. This is equivalent to removing a multiplicative factor from the light curve in flux space. To verify that we accurately removed the systematic photometric offset, we also apply the median calibration to the reference stars themselves, and calculate the $\chi^2$ relative to the mean flux of the stellar light curves before and after the calibration. We show the distribution of the $\chi^2$ of the reference stars for \obj \ in Extended Data Fig. \ref{fig:chi2}. The distribution of the calibrated stellar light curves follows the expected $\chi^2$ distribution, while the uncalibrated ones yield systematically higher $\chi^2$. This verifies that our calibration process successfully removes the systematic photometric offset from the light curves, and that the estimated photometric uncertainties satisfactorily represent the observed errors.

We then calculate a $\chi^2$ from the calibrated quasar light curves to quantify the significance of variability in the WISE bands. Among the 30 multi-epoch WISE-detected quasars, two quasars show a $\chi^2$ corresponding to a $5\sigma$ or higher significance in their variability, while six are at $3\sigma$ or higher. The quasar with the most significant variability is \obj , with a $\approx 10\sigma$ significance. 

\begin{figure}[t]
\centering
\includegraphics[width=0.9\textwidth]{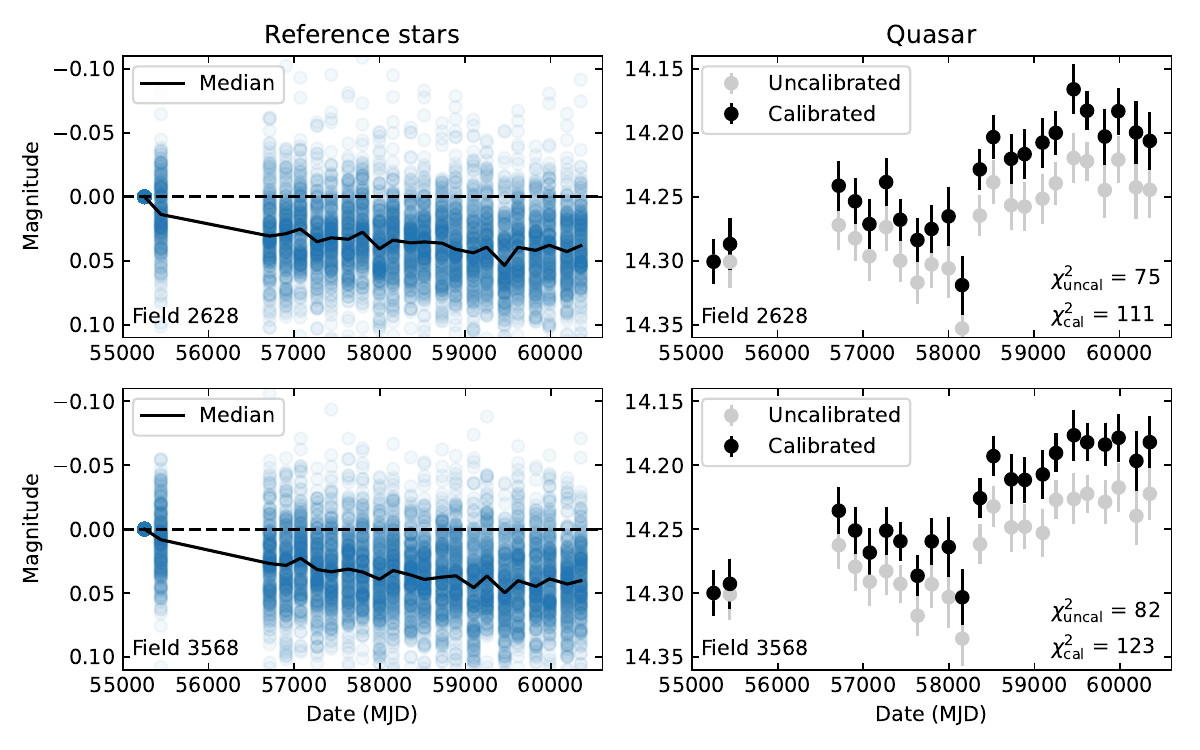}
\caption{{\bf Demonstration of the WISE light curve calibration process.} Left: The uncalibrated W1 light curves for the reference stars of \obj . The blue points show the light curves for the individual stars, while the black solid line shows the median light curve. The quasar light curve is calibrated by subtracting the median stellar light curve. Right: The W1 light curves for \obj \ before (grey) and after (black) calibration by subtracting the median stellar light curves. The top and bottom panels show the light curves in the two different WISE fields covering \obj . Error bars represent $1\sigma$ uncertainties.}\label{fig:cal}
\end{figure}

In the left panels of Extended Data Fig. \ref{fig:cal}, we show the uncalibrated light curves of the reference stars for \obj , as well as their median. \obj \ is located in the overlapping region of two WISE fields, 2628 and 3568, resulting in one set of observations per field, which are treated independently in our calibration. In both fields, the uncalibrated median indicates a decreasing measured brightness for the reference stars over time. In the right panels, of Extended Data Fig. \ref{fig:cal}, we show the light curves of \obj \ before and after the flux calibration. While the flux calibration led to a lower $\chi^2$ for the reference stars as shown in Extended Data Fig. \ref{fig:chi2}, the $\chi^2$ for \obj \ increased from $\sim 80$ to $\gtrsim 110$ in each field. This suggests that the observed variability is not due to the same systematic photometric offset seen in the reference stars.

\subsection{Ground-based data}
We incorporate archival ground-based infrared observations for \obj \ in our analysis. \obj \ is covered the UKIRT Hemisphere Survey (UHS) \citep{dye18, schneider25}, which observed the quasar in the $J$ band in 2013 and the $K$ band in 2018, each with four 10 s exposures. Additionally, \obj \ was also observed by the Two Micron All Sky Survey (2MASS) \citep{skrutskie06} in $JHK$ in 1997, with an integration time of 7.8 s per filter. We retrieve the UHS and 2MASS fluxes from their respective public catalogs. For consistency, we convert the 2MASS $J$ and $K$ magnitudes to the UHS filter system by calculating a magnitude offset, derived from the quasar's infrared spectral shape based on the GNIRS spectrum presented in \citet{fan19}.

\subsection{HST data}
\obj \ was observed with {\it HST}/WFC3 in 2020 by GO program 15829 (PI: Yang), with infrared imaging obtained in the F125W and F160W filters. We retrieved the reduced mosaics from the Mikulski Archive for Space Telescopes
(MAST; \url{https://archive.stsci.edu}), and perform a background subtraction using SEP \citep{barbary16}, a Python implementation of SExtractor \citep{bertin96}. Fluxes in each filter were measured using a 12 pixel (1.54'') radius aperture centered on the quasar position. We estimate the errors from the error map in the mosaic, which accounts for both background and Poisson noise. At the end, we converted the F125W and F160W magnitudes into the UHS $J$ band and 2MASS $H$ band, respectively.

\begin{figure}[htp]
\centering
\includegraphics[width=0.99\textwidth]{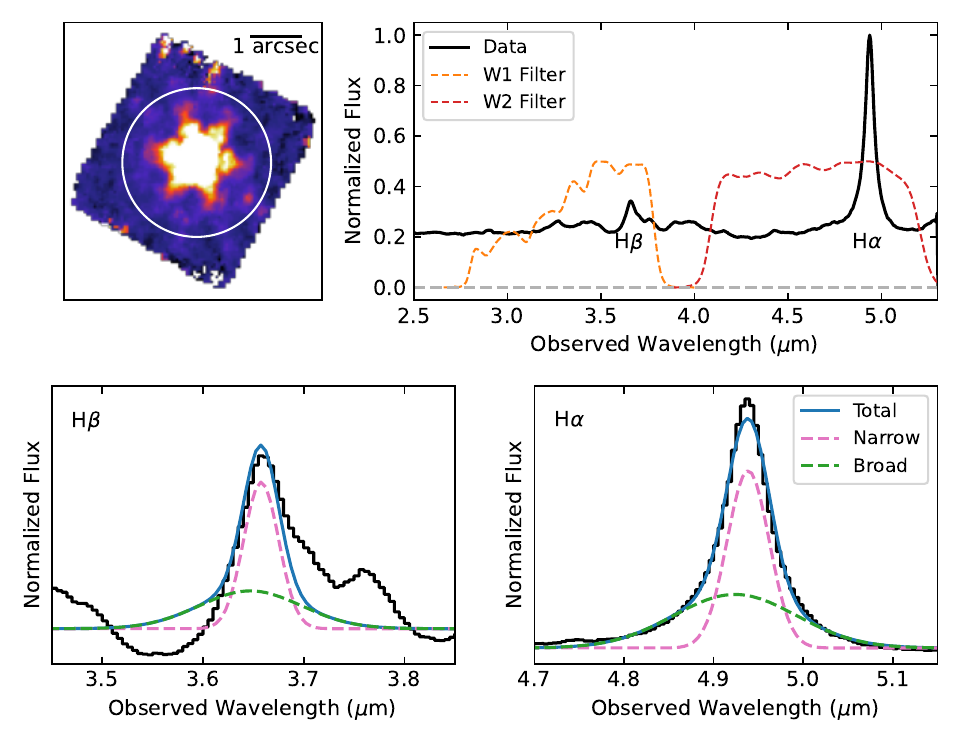}
\caption{{\bf JWST/NIRSpec IFU image and spectrum for \obj .} Left: NIRSpec IFU flux map of \obj \ at 4.5 $\mu$m. The white circle shows the 1.5'' radius aperture used to extract the spectrum. Middle: The extracted spectrum is shown in the black lines, while the WISE W1 and W2 filter transmission curves are shown in the orange and red dashed lines. Top right: The H$\beta$ emission line spectrum (black histogram), the best-fit model (blue curve), and the narrow (magenta dashed line) and broad (green dashed line) components of the fit. Bottom right: Same as the top right panel, but for the H$\alpha$ emission line. The emission from the broad H$\beta$ and H$\alpha$ components in the W1 and W2 filters is 1.9\% and 11.1\% of the continuum, respectively.}\label{fig:nirspec}
\end{figure}

\subsection{JWST data}

The JWST GTO program 1222 (PI: Willott) observed \obj \ in January 2023 using both NIRSpec high-resolution grating fixed slit spectroscopy and PRISM IFU spectroscopy. The fixed slit spectroscopy is subject to incomplete wavelength coverage over the infrared filters due to chip gaps, and slit loss due to a slight misalignment between the slit and the center of the quasar position. Thus, for this study, we utilize the IFU observations to extract photometry of the quasar. The IFU observations were obtained with the G395H/F290LP and PRISM/CLEAR grating-filter combinations. We focus on the latter for our analysis, as it covers the wavelength range corresponding to the $J$, $H$ and $K$ bands. 

We reduced the NIRSpec IFU data using the STScI JWST pipeline version 1.17.1 with Calibration Reference Data System version 12.0.9 (\texttt{jwst\_1299.pmap}). First, we run the standard infrared detector reductions with \texttt{Detector1Pipeline}. Next, we correct the \texttt{.rate} files for $1/f$ noise \citep{Schlawin2020} by using a running mean algorithm, and for snowball effects and cosmic rays using \texttt{snowblind} (\url{https://github.com/mpi-astronomy/snowblind}). Then, we run \texttt{Spec2Pipeline} to  produce calibrated spectral data and build 3D datacubes for each dither exposure. These datacubes are aligned and combined into a single datacube using the Photutils \texttt{reproject} routine \citep{Vayner2023}, after applying a sigma clip routine that masks pixels with extreme outliers. The final datacube is resampled with a spatial resolution of $0.05''$ per spaxel. Since no dedicated background exposures were taken, we perform aperture background subtraction on the final combined datacube. 

To derive photometry, we construct images in the $J$, $H$ and $K$ bands by integrating the data cube along the wavelength axis over the respective filter transmission curves. We then perform aperture photometry in a 30 pixel (1.5'') radius aperture. Photometric uncertainties are estimated from the associated error cube by propagating the variance across the integrated spectral range.

To estimate the contribution of the broad line region emission to the W1 and W2 photometry, we measure the broad H$\beta$ and H$\alpha$ line fluxes from the spectrum extracted in a 1.5'' radius aperture in the IFU data cube. We show the IFU flux map at 4.5 $\mu$m and the extraction region in the left panel of Extended Data Fig. \ref{fig:nirspec}, and the extracted spectrum in the middle panel. We model each emission line as the sum of two Gaussians, representing the narrow and broad components, and a constant continuum. The narrow and broad components have independent amplitudes, central wavelengths and velocity dispersions, but the central wavelengths for each component are tied between H$\alpha$ and H$\beta$. We obtain a posterior distribution for the parameters by Markov Chain Monte Carlo (MCMC) using the \texttt{emcee} package. The emission line profiles, best-fit models and their components for the H$\beta$ and H$\alpha$ emission lines are shown in the right panels of Extended Data Fig. \ref{fig:nirspec}. We then integrate the best-fit broad Gaussian components over the W1 and W2 filter curves to measure their flux in $f_\nu$. We find that the broad H$\beta$ and H$\alpha$ fluxes are 1.9\% and 11.1\% of the continuum level, respectively.

\subsection{eRORISTA data}

\obj \ lies within the footprint of the eROSITA (extended ROentgen Survey with an Imaging Telescope Array; \citealt{Predehl21}) All-Sky Survey \citep[eRASS;][]{Merloni24}. With the eRASS scanning strategy, eROSITA maps the entire sky every six months from L2, with the survey starting in December 2019. During each scan, a given sky position is typically visited six times for up to 40 s per visit. The instrument is most sensitive in the soft X-ray band 0.2 -- 2.3 keV (corresponding to 1.5 -- 17.3 keV in the quasar rest frame), and attains an on-axis resolution of 16.1'' half-energy width at 1.5 keV. 
Thus far, \obj \ has been covered in four consecutive eRASS scans. While no source is significantly detected in the first three passes, a source with a positional uncertainty of 4.70’' is detected within 2.78'' of \obj \ in the fourth pass (eRASS4). This detection occurred on 2021 September 4, i.e. 376.6 days in the observer frame (50.2 days in the quasar rest frame) after the deep \emph{XMM-Newton} pointing reported by \citet{Yang22}. The vignetted exposure time at the source position in the 0.2—2.3 keV band is 131 s. 
The detection likelihood, defined as the negative natural logarithm of the probability that the observed X-ray signal is due to background fluctuations, is \texttt{DET\_LIKE} $=15.5$ in the 0.2–2.3 keV detection band. This detection is based on six counts concentrated in the 1--2\,keV range. We converted the eRASS4 count rate into fluxes in the observer-frame 0.5–2 keV and 2–10 keV bands, obtaining $f_{0.5-2\,\mathrm{keV}} = (4.01 \pm 1.80) \times 10^{-14}\,\mathrm{erg\,cm^{-2}\,s^{-1}}$ and $f_{2-10\,\mathrm{keV}} = (1.22 \pm 0.55) \times 10^{-13}\,\mathrm{erg\,cm^{-2}\,s^{-1}}$, assuming a power law with photon index $\Gamma = 1.9$ and an intrinsic absorption of $2.8 \times 10^{23}\,\mathrm{cm^{-2}}$ \citep{Yang22}.

\subsection{Extinction correction}
The fluxes and uncertainties are corrected for Galactic extinction assuming $E(B-V) = 0.52$, estimated by \citet{schlafly11} based on the quasar position, and a \citet{cardelli89} Milky Way extinction curve.

\subsection{Variable SED Analysis}
In Fig. \ref{fig:varspec}, we show the variable SED, defined as the difference between the maximum and minimum fluxes as a function of wavelength. We fit the observed data with a $\lambda f_\lambda \propto \lambda^\alpha$\ model by MCMC using the \texttt{emcee} package, where we obtain the median and $68\%$ confidence interval of the model plotted in Fig. \ref{fig:varspec}. The model is strongly constrained by the data points in the $J$ and W1 bands, which have the smallest observational uncertainties. The data points in the $K$ and W2 bands deviate from the model by $\sim 1\sigma$, which is within observational uncertainties. The observed $H$ band flux is $\sim 2\sigma$ higher than the model. For this band, the minimum state is determined by the 2MASS observation in 1997, which predates the WISE observations from 2010 to 2024 and the $J$ band UHS observation in 2013 by over 10 years. As a result, the minimum state in $H$ (as well as $K$), could represent a lower state than those in the $J$ and WISE bands, leading to a larger difference between the maximum and minimum, and a larger deviation from the model. Nonetheless, the $J$ and WISE band observations, which agree with the best-fit relation, share a common coverage in $\sim 2013-2024$, and their minimum and maximum states are taken within the same period. Therefore, we consider the best-fit spectral slope robust within the period covered by the WISE observations.

\subsection{Timescale Analysis}\label{sec:timescale}
We estimate the relevant timescale for the variability \citep[see e.g.][]{burke21}. The orbital timescale is given by 
\begin{equation}
    t_\mathrm{orb} \simeq 100 \left(\frac{R}{100R_S}\right)^{3/2}\left(\frac{M_\bullet}{10^8 M_\odot}\right) \mathrm{days}.
\end{equation}
The thermal timescale is given by 
\begin{equation}
    t_\mathrm{th} \simeq 1680 \left(\frac{\alpha}{0.01}\right)^{-1} \left(\frac{R}{100R_S}\right)^{3/2}\left(\frac{M_\bullet}{10^8 M_\odot}\right) \mathrm{days}.
\end{equation}
The viscous timescale is given by
\begin{equation}
    t_\mathrm{th} \simeq \left( \frac{H}{R} \right)^{-2} t_\mathrm{th}.
\end{equation}
$M_\bullet$ is the black hole mass, $\alpha$ is the viscosity parameter, and $H$ is the disk height in the equations above. We assume $\alpha=0.03$ and $H/R=0.1$, which are common for thin disks.

In our analysis, we assume $R=200 R_S$, taken from the radius at which the temperature produces a black body spectrum peaking at the W1 wavelength.
This is calculated using the temperature profile of the thin disk model \citep{peterson97}, which gives
\begin{equation}\label{eq:temp}
    T(r) \approx 6.3 \times 10^5 \left(\frac{\dot{M}}{\dot{M_\mathrm{Edd}}} \right)^{1/4}  \left(\frac{M_\bullet}{M_\odot} \right)^{-1/4} \left(\frac{R}{R_S} \right)^{-3/4}~\mathrm{K}, 
\end{equation}
where $\dot{M}$ is the accretion rate, $\dot{M_\mathrm{Edd}}$ is the Eddington accretion rate, $M_\bullet$ is the black hole mass, and $R_S$ is the Schwarzschild radius. Using Wien's law, a blackbody spectrum at 6500 K peaks at the rest-frame wavelength corresponding to the W1 filter. Equation \ref{eq:temp} implies that the emission region responsible for this temperature is at $R/R_S = 203.5$.

\begin{figure}[t]
\centering
\includegraphics[width=0.9\textwidth]{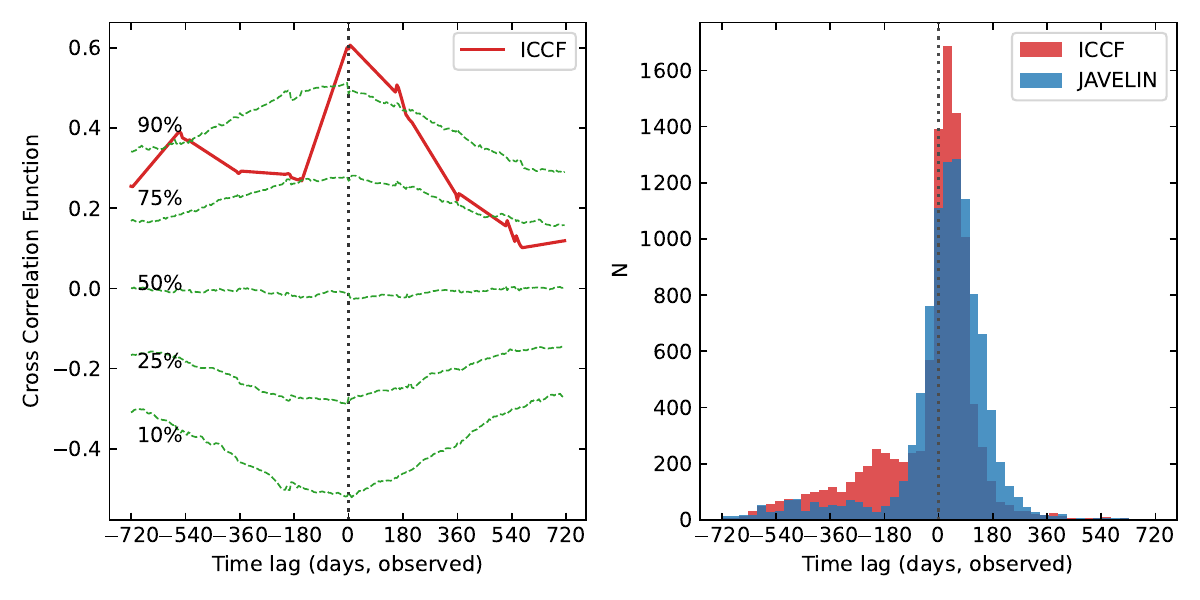}
\caption{{\bf Time lag analysis of the W1 and W2 light curves.} Left: The ICCF between the observed W1 and W2 light curves, shown in the red solid line. The green dashed lines show the 10, 25, 50, 75 and 90 percent confidence intervals from simulations. Right: The posterior distribution of the time lag. The ICCF centroids from the FR/RSS procedure are shown in red, while results from \texttt{JAVELIN} are shown in blue.}\label{fig:iccf}
\end{figure}

\subsection{Time Delay Analysis}
We measure the time lag between the W2 and W1 light curves using the interpolated cross-correlation function (ICCF) method \citep{gaskell87} and the \texttt{JAVELIN} software \citep{zu11, zu13}. For the ICCF method, one light curve is shifted and interpolated on a grid of time delays, and the correlation coefficient is calculated at each step. We then estimate the lag from the centroid of the time delay within $80\%$ of the ICCF peak. The process is repeated by shifting and interpolating the other light curve, and the final time lag is calculated by taking the average of the two. We show the ICCF for the observed light curves in the left panel of Extended Data Fig. \ref{fig:iccf}. The uncertainties and confidence intervals are estimated using the flux randomization/random subset selection (FR/RSS) method \citep{peterson98, peterson04}, calculating a distribution of ICCF from 5000 random realizations of the light curves. Fluxes are sampled by a Gaussian distribution based on the associated flux errors, and data points are randomly drawn with replacement on the light curve, discarding the repeated data points. In the right panel of Extended Data Fig. \ref{fig:iccf}, we show the distribution of the resulting ICCF centroids from the FR/RSS procedure.

To test the significance of the ICCF peak, we simulate W1 and W2 light curves based on the observed variability properties in a procedure similar to that in \citet{breedt09}. We measure the power spectral density of the observed W1 light curve from a continuous-time autoregressive moving average (CARMA) model using the \texttt{CARMA\_pack} package \citep{kelly14}. We then generate 1000 light curves for W1 and W2 based-on the measured CARMA model, perturbed by the observed flux errors in each filter. An ICCF is calculated for each realization of these pairs of uncorrelated light curves to estimate the probability of obtaining the observed ICCF peak by chance. We plot the 10th, 25th, 50th, 75th and 90th percentiles of the simulated ICCF in the left panel of Extended Data Fig. \ref{fig:iccf}. The observed ICCF peak is at approximately the 95th percentile level.

In addition to the ICCF method, we also measure the time lag using \texttt{JAVELIN}. Instead of linear interpolation, \texttt{JAVELIN} models the intrinsic light curve with a Damped Random Walk model, while the delayed light curve is offset by a time lag, smoothed by a top-hat function, and scaled by a normalization factor. We show the distribution of the posterior for the time lag in the blue histogram in the right panel of Extended Data Fig. \ref{fig:iccf}. The time lag measured from \texttt{JAVELIN} is comparable to those from the ICCF method, with a median of 40 days and a $90\%$ upper limit of 160 days, in the observed frame. This is consistent with no observable time lag within the 180-day cadence. In the main text, we report the upper limit value from \texttt{JAVELIN}.

We estimate the disk size at the W2 wavelength assuming the standard thin disk model, where the time delay is expected to follow $\tau \propto \lambda ^{4/3}$. We calculate the size of the disk where the emission peaks at the W2 wavelength by extrapolating the W2-W1 time delay to 912~\AA, which is commonly assumed to represent the true ionizing continuum of the accretion disk \citep[e.g.][]{fausnaugh16}. We then compare this to the expected disk size described in the previous subsection.

In the local universe, the time lag has been observed to follow a scaling relation with the AGN luminosity \citep[e.g.][]{netzer22, montano22}. We compare time lag our upper limit with the lag-luminosity relation reported in \citet{montano22}, which finds $\log(\tau_{gz}/\mathrm{min}) = 0.56 \log(L_{5100}/\mathrm{erg\ s^{-1}}) - 20.87$, where $\tau_{gz}$ is the time lag in the SDSS $g$ and $z$ bands. We convert $\tau_{gz}$ to the time lag between W1 and W2 ($\tau_{12}$) using the relation $\tau(\lambda) = \tau_0[(\lambda/\lambda_0)^{4/3}-1]$) expected for a thin disk. At the $L_\mathrm{5100}$ of \obj , this yields $\tau_{12} = 19.6$ days in the observed frame, which is consistent with our observed upper limit.

\subsection{Microlensing Analysis}

\setcounter{figure}{0}
\renewcommand{\thefigure}{\arabic{figure}}
\renewcommand{\figurename}{Supplementary Fig.}
\renewcommand{\theHfigure}{ext.\arabic{figure}}

\begin{figure}[t]
\centering
\includegraphics[width=0.5\textwidth]{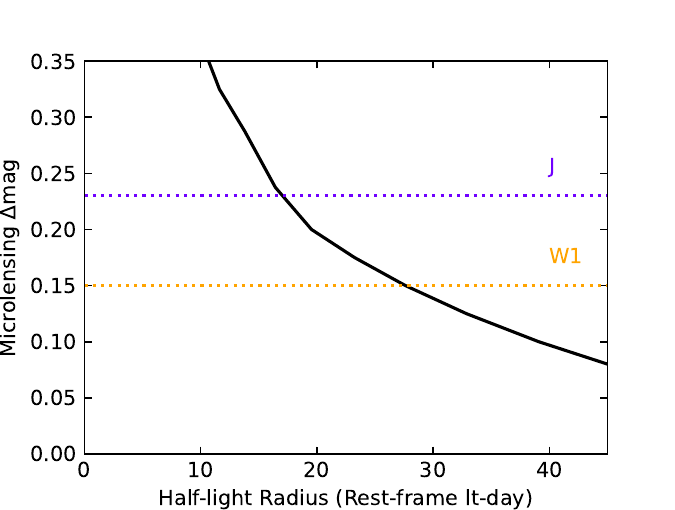}
\caption{{\bf The $95\%$ width of the microlensing magnification as a function of the source size.} The dotted lines show the observed variability amplitudes in $J$ and W1. The maximum disk size in the $J$ and W1 filters are 17 and 28 rest-frame light-days, respectively.} \label{fig:micro}
\end{figure}

Since \obj \ is known to be strongly lensed, we investigate the possibility of flux variations due to microlensing by stars within the lens galaxy. To do this, we measure the statistical distribution of magnification due to microlensing following procedures similar to \citet{Weisenbach21}, and compare that with our observed light curves.

We adopt the total convergence and shear at the quasar macroimage positions using the lens model in \citet{fan19} and M. Yue et al. [in prep.]. Assuming typical values for the stellar mass fraction and average stellar mass of $\kappa_*/\kappa = 0.1$ and $m_*=0.3\ M_\odot$ respectively, we create microlensing magnification maps for each of the multiple macroimages. By convolving the maps with Gaussian profiles of various sizes, we estimate the amount of (de)magnification that is possible as a function of source size. Since the images are not resolved in the time domain, we sum up the possible magnifications of the multiple images to obtain the total unresolved magnification probability distribution as a function of source size. 

As the source size increases, the width of the microlensing magnification distribution shrinks. Comparing the width of the distribution with the observed variability amplitude provides an estimate of the upper limits of the quasar accretion disk size in the various bands, as sources with a larger size will not be able to produce variations as large as those observed.

We show the resulting $95\%$ width of the magnification distribution as a function of the source half-light radius in Supplementary Fig. \ref{fig:micro}. The microlensing magnification widths imply an upper limit of 17 and 28 rest-frame light-days for the $J$-band and W1 emission region. These upper limits are consistent with the thin disk model predictions (Equation \ref{eq:temp}) of 3.9 and 15 rest-frame light-days for the two filters. However, the microlensing-inferred upper limits for the $J$ and W1 bands imply a radial scaling of  $R \propto \lambda^\beta$ with $\beta \approx 0.49$. This slope is significantly shallower than expectations from both the standard thin disk ($\beta=4/3$) and slim disk ($\beta = 2$) models.

We estimate the timescale of a potential microlensing event following the procedures in \citet{Mosquera11}. The timescale is given by the size of the source on the source plane divided by the effective velocity of the source,
\begin{equation}
v_{\rm eff}^2 =
\left(\frac{\sigma_{\rm pec}(z_l)}{1+z_l}\frac{D_{OS}}{D_{OL}}\right)^2 +
\left(\frac{\sigma_{\rm pec}(z_s)}{1+z_s}\right)^2 +
\left(\frac{v_{\rm CMB}}{1+z_l}\frac{D_{LS}}{D_{OL}}\right)^2 +
2\left(\frac{\sigma_*}{1+z_l}\frac{D_{OS}}{D_{OL}}\right)^2 ,
\end{equation}
where $\sigma_\mathrm{pec}$ is the one-dimensional rms galaxy peculiar velocity, $v_\mathrm{CMB}$ is the projection of the cosmic microwave background (CMB) dipole velocity onto the lens plane, $\sigma_*$ is the velocity  dispersion of the stars in the lens galaxy, $D_{OS}$, $D_{OL}$, $D_{LS}$ are the angular diameter distances between the observer, source and lens, and $z_l$ and $z_s$ are the redshifts of the lens and source.
Evaluating $\sigma_\mathrm{pec}$ at the redshifts of the lens and source using Eq. 6 of \citet{Mosquera11} yields $\sigma_\mathrm{pec}(z_l)=275~\kms$ and $\sigma_\mathrm{pec}(z_s)=126~\kms$, while Eq. 7 of \citet{Mosquera11} gives $\sigma_*=70~\kms$ with an image separation of $\approx0.2''$ for \obj \ \citep{fan19}. At the coordinates of \obj , the CMB dipole velocity is $v_\mathrm{CMB} = 364~\kms$. Taken together, this gives an effective velocity of $v_\mathrm{eff}\approx 180 ~\kms$. Assuming a source size of $\sim 10$ light days, similar to the thin disk model prediction and within the microlensing upper limits, this yields a microlensing event duration of 45 years in the observed frame, which is nearly an order of magnitude longer than the $\sim5$-year duration between the observed W1 flux minimum in 2016 and the maximum in 2021. Therefore, we conclude that it is unlikely that the observed variability in \obj \ is purely due to a microlensing event, and intrinsic disk variability remains our preferred interpretation of the observed variability.

\subsection{Cosmological Parameters}
Throughout this article, we adopt the cosmological parameters from the Planck 2018 results \citep{planck20}.

\subsection{Data availability}
The JWST, HST, UHS, 2MASS data are publicly available. The calibrated WISE light curves are provided as Supplementary Data.

\subsection{Code availability}
The JWST data were reduced using the publicly available JWST Calibration Pipeline (\url{https://jwst-pipeline.readthedocs.io/en/stable/}). The \texttt{JAVELIN} time delay analysis code is publicly available (\url{https://github.com/legolason/javelin-1}). The WISE data calibration code can be provided upon request.

\subsection{Acknowledgements}
We thank Paul Schechter and Eduardo Ba\~nados for insightful discussions. We also thank the reviewers for their constructive comments, which helped improve the quality of this paper. G.C.K.L, A.C.E., C.P. and K.D. acknowledge the support of the National Aeronautics and Space Administration through ADAP grant number 80NSSC24K0663. M.K. is supported by DLR grant FKZ 50 OR 2519.

\subsection{Author Contributions}
G.C.K.L. led the analysis and interpretation, and wrote the manuscript. A.C.E. conceived the variability search with WISE. C.P. contributed to the direction of the analysis and interpretation. X.F. and M.Y. contributed to the characterization of the quasar. K.D. reduced the WISE data. J.W., A.M. and M.K. conducted the eROSITA data analysis. Y.I. reduced the JWST/NIRSpec IFU data. J.Y. reduced the ground-based spectra. L.W. conducted the microlensing magnification analysis. M.Y. reduced the HST/WFC3 data and produced the macrolensing model. C.P., X.F., E.K., R.A.S., F.W., J.Y. assisted with the interpretation of the results. All authors contributed to the manuscript.

\subsection{Competing Interests}
The authors declare no competing interests.



\subsection{References}

\begin{thebibliography}{56}
\ifx \bisbn   \undefined \def \bisbn  #1{ISBN #1}\fi
\ifx \binits  \undefined \def \binits#1{#1}\fi
\ifx \bauthor  \undefined \def \bauthor#1{#1}\fi
\ifx \batitle  \undefined \def \batitle#1{#1}\fi
\ifx \bjtitle  \undefined \def \bjtitle#1{#1}\fi
\ifx \bvolume  \undefined \def \bvolume#1{\textbf{#1}}\fi
\ifx \byear  \undefined \def \byear#1{#1}\fi
\ifx \bissue  \undefined \def \bissue#1{#1}\fi
\ifx \bfpage  \undefined \def \bfpage#1{#1}\fi
\ifx \blpage  \undefined \def \blpage #1{#1}\fi
\ifx \burl  \undefined \def \burl#1{\textsf{#1}}\fi
\ifx \betal  \undefined \def \betal{\textit{et al.}}\fi
\ifx \binstitute  \undefined \def \binstitute#1{#1}\fi
\ifx \binstitutionaled  \undefined \def \binstitutionaled#1{#1}\fi
\ifx \bctitle  \undefined \def \bctitle#1{#1}\fi
\ifx \beditor  \undefined \def \beditor#1{#1}\fi
\ifx \bpublisher  \undefined \def \bpublisher#1{#1}\fi
\ifx \bbtitle  \undefined \def \bbtitle#1{#1}\fi
\ifx \bedition  \undefined \def \bedition#1{#1}\fi
\ifx \bseriesno  \undefined \def \bseriesno#1{#1}\fi
\ifx \blocation  \undefined \def \blocation#1{#1}\fi
\ifx \bsertitle  \undefined \def \bsertitle#1{#1}\fi
\ifx \bsnm \undefined \def \bsnm#1{#1}\fi
\ifx \bsuffix \undefined \def \bsuffix#1{#1}\fi
\ifx \bparticle \undefined \def \bparticle#1{#1}\fi
\ifx \barticle \undefined \def \barticle#1{#1}\fi
\bibcommenthead
\ifx \bconfdate \undefined \def \bconfdate #1{#1}\fi
\ifx \botherref \undefined \def \botherref #1{#1}\fi
\ifx \url \undefined \def \url#1{\textsf{#1}}\fi
\ifx \bchapter \undefined \def \bchapter#1{#1}\fi
\ifx \bbook \undefined \def \bbook#1{#1}\fi
\ifx \bcomment \undefined \def \bcomment#1{#1}\fi
\ifx \oauthor \undefined \def \oauthor#1{#1}\fi
\ifx \citeauthoryear \undefined \def \citeauthoryear#1{#1}\fi
\ifx \endbibitem  \undefined \def \endbibitem {}\fi
\ifx \bconflocation  \undefined \def \bconflocation#1{#1}\fi
\csname PreBibitemsHook\endcsname

\bibitem[\protect\citeauthoryear{{Fausnaugh} et~al.}{2016}]{fausnaugh16}
\begin{barticle}
\bauthor{\bsnm{{Fausnaugh}}, \binits{M.M.}}, \betal:
\batitle{{Space Telescope and Optical Reverberation Mapping Project. III.
  Optical Continuum Emission and Broadband Time Delays in NGC 5548}}.
\bjtitle{Astrophys. J.}
\bvolume{821}(\bissue{1}),
\bfpage{56}
(\byear{2016})
\end{barticle}
\endbibitem

\bibitem[\protect\citeauthoryear{{Cackett} et~al.}{2020}]{cackett20}
\begin{barticle}
\bauthor{\bsnm{{Cackett}}, \binits{E.M.}}, \betal:
\batitle{{Supermassive Black Holes with High Accretion Rates in Active Galactic
  Nuclei. XI. Accretion Disk Reverberation Mapping of Mrk 142}}.
\bjtitle{Astrophys. J.}
\bvolume{896}(\bissue{1}),
\bfpage{1}
(\byear{2020})
\end{barticle}
\endbibitem

\bibitem[\protect\citeauthoryear{{Pozo Nu{\~n}ez} et~al.}{2025}]{pozonunez25}
\begin{barticle}
\bauthor{\bsnm{{Pozo Nu{\~n}ez}}, \binits{F.}},
\bauthor{\bsnm{{Ba{\~n}ados}}, \binits{E.}},
\bauthor{\bsnm{{Panda}}, \binits{S.}},
\bauthor{\bsnm{{Heidt}}, \binits{J.}}:
\batitle{{Accretion disc reverberation mapping in a high-redshift quasar}}.
\bjtitle{Astron. Astrophys.}
\bvolume{700},
\bfpage{8}
(\byear{2025})
\end{barticle}
\endbibitem

\bibitem[\protect\citeauthoryear{{Vanden Berk} et~al.}{2004}]{vandenberk04}
\begin{barticle}
\bauthor{\bsnm{{Vanden Berk}}, \binits{D.E.}}, \betal:
\batitle{{The Ensemble Photometric Variability of
  \raisebox{-0.5ex}\textasciitilde25,000 Quasars in the Sloan Digital Sky
  Survey}}.
\bjtitle{Astrophys. J.}
\bvolume{601}(\bissue{2}),
\bfpage{692}--\blpage{714}
(\byear{2004})
\end{barticle}
\endbibitem

\bibitem[\protect\citeauthoryear{{MacLeod} et~al.}{2012}]{macleod12}
\begin{barticle}
\bauthor{\bsnm{{MacLeod}}, \binits{C.L.}}, \betal:
\batitle{{A Description of Quasar Variability Measured Using Repeated SDSS and
  POSS Imaging}}.
\bjtitle{Astrophys. J.}
\bvolume{753}(\bissue{2}),
\bfpage{106}
(\byear{2012})
\end{barticle}
\endbibitem

\bibitem[\protect\citeauthoryear{{Benati Gon{\c{c}}alves}
  et~al.}{2025}]{goncalves25}
\begin{barticle}
\bauthor{\bsnm{{Benati Gon{\c{c}}alves}}, \binits{H.}},
\bauthor{\bsnm{{Panda}}, \binits{S.}},
\bauthor{\bsnm{{Storchi Bergmann}}, \binits{T.}},
\bauthor{\bsnm{{Cackett}}, \binits{E.M.}},
\bauthor{\bsnm{{Eracleous}}, \binits{M.}}:
\batitle{{Exploring Quasar Variability with ZTF at $0 < z < 3$: A Universal
  Relation with the Eddington Ratio}}.
\bjtitle{Astrophys. J.}
\bvolume{988}(\bissue{1}),
\bfpage{27}
(\byear{2025})
\end{barticle}
\endbibitem

\bibitem[\protect\citeauthoryear{{York} et~al.}{2000}]{york00}
\begin{barticle}
\bauthor{\bsnm{{York}}, \binits{D.G.}}, \betal:
\batitle{{The Sloan Digital Sky Survey: Technical Summary}}.
\bjtitle{Astron. J.}
\bvolume{120}(\bissue{3}),
\bfpage{1579}--\blpage{1587}
(\byear{2000})
\end{barticle}
\endbibitem

\bibitem[\protect\citeauthoryear{{Bellm} et~al.}{2019}]{bellm19}
\begin{barticle}
\bauthor{\bsnm{{Bellm}}, \binits{E.C.}}, \betal:
\batitle{{The Zwicky Transient Facility: System Overview, Performance, and
  First Results}}.
\bjtitle{Publ. Astron. Soc. Pac.}
\bvolume{131}(\bissue{995}),
\bfpage{018002}
(\byear{2019})
\end{barticle}
\endbibitem

\bibitem[\protect\citeauthoryear{{Fan} et~al.}{2023}]{fan23}
\begin{barticle}
\bauthor{\bsnm{{Fan}}, \binits{X.}},
\bauthor{\bsnm{{Ba{\~n}ados}}, \binits{E.}},
\bauthor{\bsnm{{Simcoe}}, \binits{R.A.}}:
\batitle{{Quasars and the Intergalactic Medium at Cosmic Dawn}}.
\bjtitle{Annu. Rev. Astron. Astrophys.}
\bvolume{61},
\bfpage{373}--\blpage{426}
(\byear{2023})
\end{barticle}
\endbibitem

\bibitem[\protect\citeauthoryear{{Abramowicz} et~al.}{1988}]{abramowicz88}
\begin{barticle}
\bauthor{\bsnm{{Abramowicz}}, \binits{M.A.}},
\bauthor{\bsnm{{Czerny}}, \binits{B.}},
\bauthor{\bsnm{{Lasota}}, \binits{J.P.}},
\bauthor{\bsnm{{Szuszkiewicz}}, \binits{E.}}:
\batitle{{Slim Accretion Disks}}.
\bjtitle{Astrophys. J.}
\bvolume{332},
\bfpage{646}
(\byear{1988})
\end{barticle}
\endbibitem

\bibitem[\protect\citeauthoryear{{Wang} et~al.}{1999}]{wang99}
\begin{barticle}
\bauthor{\bsnm{{Wang}}, \binits{J.-M.}},
\bauthor{\bsnm{{Szuszkiewicz}}, \binits{E.}},
\bauthor{\bsnm{{Lu}}, \binits{F.-J.}},
\bauthor{\bsnm{{Zhou}}, \binits{Y.-Y.}}:
\batitle{{Emergent Spectra from Slim Accretion Disks in Active Galactic
  Nuclei}}.
\bjtitle{Astrophys. J.}
\bvolume{522}(\bissue{2}),
\bfpage{839}--\blpage{845}
(\byear{1999})
\end{barticle}
\endbibitem

\bibitem[\protect\citeauthoryear{{Nakajima} et~al.}{2023}]{nakajima23}
\begin{barticle}
\bauthor{\bsnm{{Nakajima}}, \binits{K.}}, \betal:
\batitle{{JWST Census for the Mass-Metallicity Star Formation Relations at z =
  4-10 with Self-consistent Flux Calibration and Proper Metallicity
  Calibrators}}.
\bjtitle{Astrophys. J. Suppl. Ser.}
\bvolume{269}(\bissue{2}),
\bfpage{33}
(\byear{2023})
\end{barticle}
\endbibitem

\bibitem[\protect\citeauthoryear{{Curti} et~al.}{2024}]{curti24}
\begin{barticle}
\bauthor{\bsnm{{Curti}}, \binits{M.}}, \betal:
\batitle{{JADES: Insights into the low-mass end of the mass-metallicity-SFR
  relation at $3 < z < 10$ from deep JWST/NIRSpec spectroscopy}}.
\bjtitle{Astron. Astrophys.}
\bvolume{684},
\bfpage{75}
(\byear{2024})
\end{barticle}
\endbibitem

\bibitem[\protect\citeauthoryear{{Jiang} et~al.}{2016}]{jiang16}
\begin{barticle}
\bauthor{\bsnm{{Jiang}}, \binits{Y.-F.}},
\bauthor{\bsnm{{Davis}}, \binits{S.W.}},
\bauthor{\bsnm{{Stone}}, \binits{J.M.}}:
\batitle{{Iron Opacity Bump Changes the Stability and Structure of Accretion
  Disks in Active Galactic Nuclei}}.
\bjtitle{Astrophys. J.}
\bvolume{827}(\bissue{1}),
\bfpage{10}
(\byear{2016})
\end{barticle}
\endbibitem

\bibitem[\protect\citeauthoryear{{Wright} et~al.}{2010}]{wright10}
\begin{barticle}
\bauthor{\bsnm{{Wright}}, \binits{E.L.}}, \betal:
\batitle{{The Wide-field Infrared Survey Explorer (WISE): Mission Description
  and Initial On-orbit Performance}}.
\bjtitle{Astron. J.}
\bvolume{140}(\bissue{6}),
\bfpage{1868}--\blpage{1881}
(\byear{2010})
\end{barticle}
\endbibitem

\bibitem[\protect\citeauthoryear{{Fan} et~al.}{2019}]{fan19}
\begin{barticle}
\bauthor{\bsnm{{Fan}}, \binits{X.}}, \betal:
\batitle{{The Discovery of a Gravitationally Lensed Quasar at z = 6.51}}.
\bjtitle{Astrophys. J. Lett.}
\bvolume{870}(\bissue{2}),
\bfpage{11}
(\byear{2019})
\end{barticle}
\endbibitem

\bibitem[\protect\citeauthoryear{{Yang} et~al.}{2021}]{yang21}
\begin{barticle}
\bauthor{\bsnm{{Yang}}, \binits{J.}}, \betal:
\batitle{{Probing Early Supermassive Black Hole Growth and Quasar Evolution
  with Near-infrared Spectroscopy of 37 Reionization-era Quasars at $6.3 < z
  {\ensuremath{\leq}} 7.64$}}.
\bjtitle{Astrophys. J.}
\bvolume{923}(\bissue{2}),
\bfpage{262}
(\byear{2021})
\end{barticle}
\endbibitem

\bibitem[\protect\citeauthoryear{{Meisner} et~al.}{2022}]{meisner22}
\begin{barticle}
\bauthor{\bsnm{{Meisner}}, \binits{A.M.}},
\bauthor{\bsnm{{Lang}}, \binits{D.}},
\bauthor{\bsnm{{Schlafly}}, \binits{E.F.}},
\bauthor{\bsnm{{Schlegel}}, \binits{D.J.}}:
\batitle{{Eight-year Full-depth unWISE Coadds}}.
\bjtitle{Res. Notes Am. Astron. Soc.}
\bvolume{6}(\bissue{3}),
\bfpage{62}
(\byear{2022})
\end{barticle}
\endbibitem

\bibitem[\protect\citeauthoryear{{Skrutskie} et~al.}{2006}]{skrutskie06}
\begin{barticle}
\bauthor{\bsnm{{Skrutskie}}, \binits{M.F.}}, \betal:
\batitle{{The Two Micron All Sky Survey (2MASS)}}.
\bjtitle{Astron. J.}
\bvolume{131}(\bissue{2}),
\bfpage{1163}--\blpage{1183}
(\byear{2006})
\end{barticle}
\endbibitem

\bibitem[\protect\citeauthoryear{{Dye} et~al.}{2018}]{dye18}
\begin{barticle}
\bauthor{\bsnm{{Dye}}, \binits{S.}}, \betal:
\batitle{{The UKIRT Hemisphere Survey: definition and J-band data release}}.
\bjtitle{Mon. Not. R. Astron. Soc.}
\bvolume{473}(\bissue{4}),
\bfpage{5113}--\blpage{5125}
(\byear{2018})
\end{barticle}
\endbibitem

\bibitem[\protect\citeauthoryear{{Schneider} et~al.}{2025}]{schneider25}
\begin{barticle}
\bauthor{\bsnm{{Schneider}}, \binits{A.C.}}, \betal:
\batitle{{The Second and Third Data Releases from the UKIRT Hemisphere
  Survey}}.
\bjtitle{Astron. J.}
\bvolume{170}(\bissue{2}),
\bfpage{86}
(\byear{2025})
\end{barticle}
\endbibitem

\bibitem[\protect\citeauthoryear{{Yang} et~al.}{2022}]{Yang22}
\begin{barticle}
\bauthor{\bsnm{{Yang}}, \binits{J.}}, \betal:
\batitle{{Deep XMM-Newton Observations of an X-ray Weak Broad Absorption Line
  Quasar at z = 6.5}}.
\bjtitle{Astrophys. J. Lett.}
\bvolume{924}(\bissue{2}),
\bfpage{25}
(\byear{2022})
\end{barticle}
\endbibitem

\bibitem[\protect\citeauthoryear{{Kara} and {Garc{\'\i}a}}{2025}]{kara25}
\begin{barticle}
\bauthor{\bsnm{{Kara}}, \binits{E.}},
\bauthor{\bsnm{{Garc{\'\i}a}}, \binits{J.}}:
\batitle{{Supermassive Black Holes in X-Rays: From Standard Accretion to
  Extreme Transients}}.
\bjtitle{Annu. Rev. Astron. Astrophys.}
\bvolume{63}(\bissue{1}),
\bfpage{379}--\blpage{430}
(\byear{2025})
\end{barticle}
\endbibitem

\bibitem[\protect\citeauthoryear{{Cackett} et~al.}{2007}]{cackett07}
\begin{barticle}
\bauthor{\bsnm{{Cackett}}, \binits{E.M.}},
\bauthor{\bsnm{{Horne}}, \binits{K.}},
\bauthor{\bsnm{{Winkler}}, \binits{H.}}:
\batitle{{Testing thermal reprocessing in active galactic nuclei accretion
  discs}}.
\bjtitle{Mon. Not. R. Astron. Soc.}
\bvolume{380}(\bissue{2}),
\bfpage{669}--\blpage{682}
(\byear{2007})
\end{barticle}
\endbibitem

\bibitem[\protect\citeauthoryear{{Kelly} et~al.}{2009}]{kelly09}
\begin{barticle}
\bauthor{\bsnm{{Kelly}}, \binits{B.C.}},
\bauthor{\bsnm{{Bechtold}}, \binits{J.}},
\bauthor{\bsnm{{Siemiginowska}}, \binits{A.}}:
\batitle{{Are the Variations in Quasar Optical Flux Driven by Thermal
  Fluctuations?}}
\bjtitle{Astrophys. J.}
\bvolume{698}(\bissue{1}),
\bfpage{895}--\blpage{910}
(\byear{2009})
\end{barticle}
\endbibitem

\bibitem[\protect\citeauthoryear{{Netzer}}{2013}]{netzer13}
\begin{bbook}
\bauthor{\bsnm{{Netzer}}, \binits{H.}}:
\bbtitle{The Physics and Evolution of Active Galactic Nuclei}.
\bpublisher{Cambridge Univ. Press},
\blocation{Cambridge}
(\byear{2013})
\end{bbook}
\endbibitem

\bibitem[\protect\citeauthoryear{{Shakura} and {Sunyaev}}{1973}]{shakura73}
\begin{barticle}
\bauthor{\bsnm{{Shakura}}, \binits{N.I.}},
\bauthor{\bsnm{{Sunyaev}}, \binits{R.A.}}:
\batitle{{Black holes in binary systems. Observational appearance.}}
\bjtitle{Astron. Astrophys.}
\bvolume{24},
\bfpage{337}--\blpage{355}
(\byear{1973})
\end{barticle}
\endbibitem

\bibitem[\protect\citeauthoryear{{Edelson} et~al.}{2015}]{edelson15}
\begin{barticle}
\bauthor{\bsnm{{Edelson}}, \binits{R.}}, \betal:
\batitle{{Space Telescope and Optical Reverberation Mapping Project. II. Swift
  and HST Reverberation Mapping of the Accretion Disk of NGC 5548}}.
\bjtitle{Astrophys. J.}
\bvolume{806}(\bissue{1}),
\bfpage{129}
(\byear{2015})
\end{barticle}
\endbibitem

\bibitem[\protect\citeauthoryear{{Abuter} et~al.}{2024}]{abuter24}
\begin{barticle}
\bauthor{\bsnm{{Abuter}}, \binits{R.}}, \betal:
\batitle{{A dynamical measure of the black hole mass in a quasar 11 billion
  years ago}}.
\bjtitle{Nature}
\bvolume{627}(\bissue{8003}),
\bfpage{281}--\blpage{285}
(\byear{2024})
\end{barticle}
\endbibitem

\bibitem[\protect\citeauthoryear{{Greene} and {Ho}}{2005}]{greene05}
\begin{barticle}
\bauthor{\bsnm{{Greene}}, \binits{J.E.}},
\bauthor{\bsnm{{Ho}}, \binits{L.C.}}:
\batitle{{Estimating Black Hole Masses in Active Galaxies Using the
  H{\ensuremath{\alpha}} Emission Line}}.
\bjtitle{Astrophys. J.}
\bvolume{630}(\bissue{1}),
\bfpage{122}--\blpage{129}
(\byear{2005})
\end{barticle}
\endbibitem

\bibitem[\protect\citeauthoryear{{Kaspi} et~al.}{2000}]{kaspi00}
\begin{barticle}
\bauthor{\bsnm{{Kaspi}}, \binits{S.}}, \betal:
\batitle{{Reverberation Measurements for 17 Quasars and the
  Size-Mass-Luminosity Relations in Active Galactic Nuclei}}.
\bjtitle{Astrophys. J.}
\bvolume{533}(\bissue{2}),
\bfpage{631}--\blpage{649}
(\byear{2000})
\end{barticle}
\endbibitem

\bibitem[\protect\citeauthoryear{{Panda} et~al.}{2024}]{panda24}
\begin{barticle}
\bauthor{\bsnm{{Panda}}, \binits{S.}},
\bauthor{\bsnm{{Pozo Nu{\~n}ez}}, \binits{F.}},
\bauthor{\bsnm{{Ba{\~n}ados}}, \binits{E.}},
\bauthor{\bsnm{{Heidt}}, \binits{J.}}:
\batitle{{Probing the C IV Continuum Size{\textendash}Luminosity Relation in
  Active Galactic Nuclei with Photometric Reverberation Mapping}}.
\bjtitle{Astrophys. J. Lett.}
\bvolume{968}(\bissue{2}),
\bfpage{16}
(\byear{2024})
\end{barticle}
\endbibitem

\bibitem[\protect\citeauthoryear{{Schlafly} et~al.}{2019}]{schlafly19}
\begin{barticle}
\bauthor{\bsnm{{Schlafly}}, \binits{E.F.}},
\bauthor{\bsnm{{Meisner}}, \binits{A.M.}},
\bauthor{\bsnm{{Green}}, \binits{G.M.}}:
\batitle{{The unWISE Catalog: Two Billion Infrared Sources from Five Years of
  WISE Imaging}}.
\bjtitle{Astrophys. J. Suppl. Ser.}
\bvolume{240}(\bissue{2}),
\bfpage{30}
(\byear{2019})
\end{barticle}
\endbibitem

\bibitem[\protect\citeauthoryear{{Lang}}{2014}]{lang14}
\begin{barticle}
\bauthor{\bsnm{{Lang}}, \binits{D.}}:
\batitle{{unWISE: Unblurred Coadds of the WISE Imaging}}.
\bjtitle{Astron. J.}
\bvolume{147}(\bissue{5}),
\bfpage{108}
(\byear{2014})
\end{barticle}
\endbibitem

\bibitem[\protect\citeauthoryear{{Barbary}}{2016}]{barbary16}
\begin{barticle}
\bauthor{\bsnm{{Barbary}}, \binits{K.}}:
\batitle{{SEP: Source Extractor as a library}}.
\bjtitle{J. Open Source Softw.}
\bvolume{1}(\bissue{6}),
\bfpage{58}
(\byear{2016})
\end{barticle}
\endbibitem

\bibitem[\protect\citeauthoryear{{Bertin} and {Arnouts}}{1996}]{bertin96}
\begin{barticle}
\bauthor{\bsnm{{Bertin}}, \binits{E.}},
\bauthor{\bsnm{{Arnouts}}, \binits{S.}}:
\batitle{{SExtractor: Software for source extraction.}}
\bjtitle{Astron. Astrophys. Suppl.}
\bvolume{117},
\bfpage{393}--\blpage{404}
(\byear{1996})
\end{barticle}
\endbibitem

\bibitem[\protect\citeauthoryear{{Schlawin} et~al.}{2020}]{Schlawin2020}
\begin{barticle}
\bauthor{\bsnm{{Schlawin}}, \binits{E.}}, \betal:
\batitle{Jwst noise floor. i. random error sources in jwst nircam time series}.
\bjtitle{Astron. J.}
\bvolume{160},
\bfpage{231}
(\byear{2020})
\end{barticle}
\endbibitem

\bibitem[\protect\citeauthoryear{{Vayner} et~al.}{2023}]{Vayner2023}
\begin{barticle}
\bauthor{\bsnm{{Vayner}}, \binits{A.}}, \betal:
\batitle{First results from the jwst early release science program q3d:
  Ionization cone, clumpy star formation, and shocks in a z = 3 extremely red
  quasar host}.
\bjtitle{Astrophys. J.}
\bvolume{955},
\bfpage{92}
(\byear{2023})
\end{barticle}
\endbibitem

\bibitem[\protect\citeauthoryear{{Predehl} et~al.}{2021}]{Predehl21}
\begin{botherref}
\oauthor{\bsnm{{Predehl}}, \binits{P.}}, et al.:
{The eROSITA X-ray telescope on SRG}.
Astron. Astrophys.,
1
(2021)
\end{botherref}
\endbibitem

\bibitem[\protect\citeauthoryear{{Merloni} et~al.}{2024}]{Merloni24}
\begin{barticle}
\bauthor{\bsnm{{Merloni}}, \binits{A.}}, \betal:
\batitle{{The SRG/eROSITA all-sky survey. First X-ray catalogues and data
  release of the western Galactic hemisphere}}.
\bjtitle{Astron. Astrophys.}
\bvolume{682},
\bfpage{34}
(\byear{2024})
\end{barticle}
\endbibitem

\bibitem[\protect\citeauthoryear{{Schlafly} and
  {Finkbeiner}}{2011}]{schlafly11}
\begin{barticle}
\bauthor{\bsnm{{Schlafly}}, \binits{E.F.}},
\bauthor{\bsnm{{Finkbeiner}}, \binits{D.P.}}:
\batitle{{Measuring Reddening with Sloan Digital Sky Survey Stellar Spectra and
  Recalibrating SFD}}.
\bjtitle{Astrophys. J.}
\bvolume{737}(\bissue{2}),
\bfpage{103}
(\byear{2011})
\end{barticle}
\endbibitem

\bibitem[\protect\citeauthoryear{{Cardelli} et~al.}{1989}]{cardelli89}
\begin{barticle}
\bauthor{\bsnm{{Cardelli}}, \binits{J.A.}},
\bauthor{\bsnm{{Clayton}}, \binits{G.C.}},
\bauthor{\bsnm{{Mathis}}, \binits{J.S.}}:
\batitle{{The Relationship between Infrared, Optical, and Ultraviolet
  Extinction}}.
\bjtitle{Astrophys. J.}
\bvolume{345},
\bfpage{245}
(\byear{1989})
\end{barticle}
\endbibitem

\bibitem[\protect\citeauthoryear{{Burke} et~al.}{2021}]{burke21}
\begin{barticle}
\bauthor{\bsnm{{Burke}}, \binits{C.J.}}, \betal:
\batitle{{A characteristic optical variability time scale in astrophysical
  accretion disks}}.
\bjtitle{Science}
\bvolume{373}(\bissue{6556}),
\bfpage{789}--\blpage{792}
(\byear{2021})
\end{barticle}
\endbibitem

\bibitem[\protect\citeauthoryear{{Peterson}}{1997}]{peterson97}
\begin{bbook}
\bauthor{\bsnm{{Peterson}}, \binits{B.M.}}:
\bbtitle{An Introduction to Active Galactic Nuclei}.
\bpublisher{Cambridge Univ. Press},
\blocation{Cambridge}
(\byear{1997})
\end{bbook}
\endbibitem

\bibitem[\protect\citeauthoryear{{Gaskell} and {Peterson}}{1987}]{gaskell87}
\begin{barticle}
\bauthor{\bsnm{{Gaskell}}, \binits{C.M.}},
\bauthor{\bsnm{{Peterson}}, \binits{B.M.}}:
\batitle{{The Accuracy of Cross-Correlation Estimates of Quasar Emission-Line
  Region Sizes}}.
\bjtitle{Astrophys. J. Suppl. Ser.}
\bvolume{65},
\bfpage{1}
(\byear{1987})
\end{barticle}
\endbibitem

\bibitem[\protect\citeauthoryear{{Peterson} et~al.}{1998}]{peterson98}
\begin{barticle}
\bauthor{\bsnm{{Peterson}}, \binits{B.M.}}, \betal:
\batitle{{On Uncertainties in Cross-Correlation Lags and the Reality of
  Wavelength-dependent Continuum Lags in Active Galactic Nuclei}}.
\bjtitle{Publ. Astron. Soc. Pac.}
\bvolume{110}(\bissue{748}),
\bfpage{660}--\blpage{670}
(\byear{1998})
\end{barticle}
\endbibitem

\bibitem[\protect\citeauthoryear{{Peterson} et~al.}{2004}]{peterson04}
\begin{barticle}
\bauthor{\bsnm{{Peterson}}, \binits{B.M.}}, \betal:
\batitle{{Central Masses and Broad-Line Region Sizes of Active Galactic Nuclei.
  II. A Homogeneous Analysis of a Large Reverberation-Mapping Database}}.
\bjtitle{Astrophys. J.}
\bvolume{613}(\bissue{2}),
\bfpage{682}--\blpage{699}
(\byear{2004})
\end{barticle}
\endbibitem

\bibitem[\protect\citeauthoryear{{Breedt} et~al.}{2009}]{breedt09}
\begin{barticle}
\bauthor{\bsnm{{Breedt}}, \binits{E.}}, \betal:
\batitle{{Long-term optical and X-ray variability of the Seyfert galaxy
  Markarian 79}}.
\bjtitle{Mon. Not. R. Astron. Soc.}
\bvolume{394}(\bissue{1}),
\bfpage{427}--\blpage{437}
(\byear{2009})
\end{barticle}
\endbibitem

\bibitem[\protect\citeauthoryear{{Kelly} et~al.}{2014}]{kelly14}
\begin{barticle}
\bauthor{\bsnm{{Kelly}}, \binits{B.C.}},
\bauthor{\bsnm{{Becker}}, \binits{A.C.}},
\bauthor{\bsnm{{Sobolewska}}, \binits{M.}},
\bauthor{\bsnm{{Siemiginowska}}, \binits{A.}},
\bauthor{\bsnm{{Uttley}}, \binits{P.}}:
\batitle{{Flexible and Scalable Methods for Quantifying Stochastic Variability
  in the Era of Massive Time-domain Astronomical Data Sets}}.
\bjtitle{Astrophys. J.}
\bvolume{788}(\bissue{1}),
\bfpage{33}
(\byear{2014})
\end{barticle}
\endbibitem

\bibitem[\protect\citeauthoryear{{Zu} et~al.}{2011}]{zu11}
\begin{barticle}
\bauthor{\bsnm{{Zu}}, \binits{Y.}},
\bauthor{\bsnm{{Kochanek}}, \binits{C.S.}},
\bauthor{\bsnm{{Peterson}}, \binits{B.M.}}:
\batitle{{An Alternative Approach to Measuring Reverberation Lags in Active
  Galactic Nuclei}}.
\bjtitle{Astrophys. J.}
\bvolume{735}(\bissue{2}),
\bfpage{80}
(\byear{2011})
\end{barticle}
\endbibitem

\bibitem[\protect\citeauthoryear{{Zu} et~al.}{2013}]{zu13}
\begin{barticle}
\bauthor{\bsnm{{Zu}}, \binits{Y.}},
\bauthor{\bsnm{{Kochanek}}, \binits{C.S.}},
\bauthor{\bsnm{{Koz{\l}owski}}, \binits{S.}},
\bauthor{\bsnm{{Udalski}}, \binits{A.}}:
\batitle{{Is Quasar Optical Variability a Damped Random Walk?}}
\bjtitle{Astrophys. J.}
\bvolume{765}(\bissue{2}),
\bfpage{106}
(\byear{2013})
\end{barticle}
\endbibitem

\bibitem[\protect\citeauthoryear{{Netzer}}{2022}]{netzer22}
\begin{barticle}
\bauthor{\bsnm{{Netzer}}, \binits{H.}}:
\batitle{{Continuum reverberation mapping and a new lag-luminosity relationship
  for AGN}}.
\bjtitle{Mon. Not. R. Astron. Soc.}
\bvolume{509}(\bissue{2}),
\bfpage{2637}--\blpage{2646}
(\byear{2022})
\end{barticle}
\endbibitem

\bibitem[\protect\citeauthoryear{{Montano} et~al.}{2022}]{montano22}
\begin{barticle}
\bauthor{\bsnm{{Montano}}, \binits{J.W.}}, \betal:
\batitle{{Optical Continuum Reverberation in the Dwarf Seyfert Nucleus of NGC
  4395}}.
\bjtitle{Astrophys. J. Lett.}
\bvolume{934}(\bissue{2}),
\bfpage{37}
(\byear{2022})
\end{barticle}
\endbibitem

\bibitem[\protect\citeauthoryear{{Weisenbach} et~al.}{2021}]{Weisenbach21}
\begin{barticle}
\bauthor{\bsnm{{Weisenbach}}, \binits{L.}},
\bauthor{\bsnm{{Schechter}}, \binits{P.L.}},
\bauthor{\bsnm{{Pontula}}, \binits{S.}}:
\batitle{{``Worst-case'' Microlensing in the Identification and Modeling of
  Lensed Quasars}}.
\bjtitle{Astrophys. J.}
\bvolume{922}(\bissue{1}),
\bfpage{70}
(\byear{2021})
\end{barticle}
\endbibitem

\bibitem[\protect\citeauthoryear{{Mosquera} and {Kochanek}}{2011}]{Mosquera11}
\begin{barticle}
\bauthor{\bsnm{{Mosquera}}, \binits{A.M.}},
\bauthor{\bsnm{{Kochanek}}, \binits{C.S.}}:
\batitle{{The Microlensing Properties of a Sample of 87 Lensed Quasars}}.
\bjtitle{Astrophys. J.}
\bvolume{738}(\bissue{1}),
\bfpage{96}
(\byear{2011})
\end{barticle}
\endbibitem

\bibitem[\protect\citeauthoryear{{Planck Collaboration}
  et~al.}{2020}]{planck20}
\begin{barticle}
\bauthor{\bsnm{{Planck Collaboration}}}, \betal:
\batitle{{Planck 2018 results. VI. Cosmological parameters}}.
\bjtitle{Astron. Astrophys.}
\bvolume{641},
\bfpage{6}
(\byear{2020})
\end{barticle}
\endbibitem

\end{thebibliography}
\end{document}